\documentstyle[aps,prd,preprint]{revtex}
\tighten
\begin{document}
\draft

\preprint{\vbox{\baselineskip=12pt
\rightline{gr-qc/9412072}
\rightline{Submitted to Physical Review D}}}

\title{The Evolution of Distorted Rotating Black Holes I: Methods and Tests}

\author{Steven R. Brandt${}^{(1,2)}$
 and Edward Seidel${}^{(1,2)}$}

\address{${}^{(1)}$ National Center for Supercomputing Applications \\
605 E. Springfield Ave., Champaign, Il. 61820}

\address{${}^{(2)}$ Department of Physics,
University of Illinois at Urbana-Champaign}

\date{\today}
\maketitle

\begin{abstract}
We have developed a new numerical code to study the evolution of
distorted, rotating black holes.  We discuss the numerical methods and
gauge conditions we developed to evolve such spacetimes.  The code has
been put through a series of tests, and we report on (a) results of
comparisons with codes designed to evolve non-rotating holes, (b)
evolution of Kerr spacetimes for which analytic properties are known,
and (c) the evolution of distorted rotating holes.  The code
accurately reproduces results of the previous NCSA non-rotating code
and passes convergence tests.  New features of the evolution of
rotating black holes not seen in non-rotating holes are identified.
With this code we can evolve rotating black holes up to about
$t=100M$, depending on the resolution and angular momentum. We also
describe a new family of black hole initial data sets which represent
rotating holes with a wide range of distortion parameters, and
distorted non-rotating black holes with odd-parity radiation.
Finally, we study the limiting slices for a maximally sliced rotating
black hole and find good agreement with theoretical predictions.
\end{abstract}


\narrowtext

\section{Introduction}
\label{intro}

In the last few years there has been much study of black holes.  They
are expected to be an important source of gravitational waves.  The
LIGO and VIRGO interferometers should begin taking data later this
decade~\cite{LIGO3}, and sensitive bar detectors are already on line
and will improve~\cite{Johnson93}.  At the same time computer power is
increasing dramatically, permitting well resolved numerical
simulations of axisymmetric black holes and the first 3D simulations
of black holes.  Collectively this work will lead ultimately to
numerical simulations of 3D spiraling, coalescing black holes, which
will be essential to interpreting the gravitational waveforms expected
to be detected.

Astrophysical black holes in the universe are expected to possess
angular momentum, yet due to the difficulty of black hole simulations
most numerical calculations of vacuum black hole initial data sets,
such as those discovered by Bowen and York~\cite{Bowen80} nearly 15
years ago, have not been attempted until now. (Recently, however,
rotating matter configurations have been successfully collapsed to the
point where a horizon forms~\cite{Abrahams94a}.)  Rotation adds a new
degree of freedom in the system, complicating matters significantly.
Even in the stationary case, the Kerr solution is much more complex
than the Schwarzschild solution.  As all black hole systems with net
angular momentum must eventually settle down to a perturbed Kerr
spacetime, it is essential to develop techniques to study distorted
rotating black holes numerically.

Such studies are interesting not only because they allow one to
examine the nonlinear evolution of distorted, rotating single black
holes, but also because they should be useful in understanding the
intermediate and late coalescence phase of the general collision of two
rotating black holes, as both cases correspond to highly distorted
Kerr spacetimes.  This parallel between distorted single black holes
and the collision of two black holes was striking in the non-rotating
case~\cite{Abrahams92a,Anninos93b,Anninos94b}, and we expect the same
to be true for the rotating case.

In this series of papers we show how one can construct and evolve
vacuum, distorted, rotating black holes.  In this first paper in this
series we present details of a numerical code designed to evolve
rotating black hole initial data sets, such as Kerr, Bowen and
York~\cite{Bowen80}, and a new class of data sets we have
developed~\cite{Brandt94a}.  With such a code we can now study the
dynamics of highly distorted, rotating black holes, paralleling recent
work of the NCSA group on non-rotating
holes~\cite{Anninos93c,Bernstein93b}.  The first studies of the
physics of these evolving systems are presented in a companion
paper~\cite{Brandt94c}, referred to henceforth as Paper II, and a
complete study of the initial data sets will be discussed in
Ref.~\cite{Brandt94a}, which we refer to as Paper III.

The paper is organized as follows:  In section~\ref{preliminaries} we
review the basic formulation of the equations, discuss our choice of
variables, and analyze symmetries present in our spacetimes.  The
initial data sets we have used are discussed in section~\ref{initdat},
and the numerical code and gauge choices are described in
section~\ref{code}.
In section~\ref{metric functions} we describe the
behavior of the metric functions in a rotating code and discuss the
differences with the non-rotating code.

\section{Preliminaries}
\label{preliminaries}
\subsection{3+1 Formalism}

We use the well known ADM~\cite{Arnowitt62} formulation of the
Einstein equations as the basis for our numerical code.  Pertinent
details are summarized here, but we refer the reader to~\cite{York79}
for a complete treatment of this formalism.  In the ADM formalism
spacetime is foliated into a set of non-intersecting spacelike
surfaces.  There are two kinematic variables which describe the
evolution between these surfaces: the lapse $\alpha$, which describes
the rate of advance of time along a timelike unit vector $n^\mu$ normal to a
surface, and the spacelike shift vector $\beta^\mu$ that describes the
motion of coordinates within a surface.  The choice of lapse and shift
is essentially arbitrary, and our choices will be described in
section~\ref{gauge} below.

The line element is written as
\begin{equation}
 ds^2   = -(\alpha^{2} -\beta _{a}\beta ^{a}) dt^2
 + 2 \beta _{a}  dx^{a} dt
 +\gamma_{ab} dx^{a} dx^{b},
\end{equation}
where $\gamma_{ab}$ is the 3--metric induced on each spacelike slice.
Given a choice of lapse $\alpha$ and shift vector $\beta^{b}$, the
Einstein equations in 3+1 formalism split into evolution equations for
the 3--metric $\gamma_{ab}$ and constraint equations that must be
satisfied on any time slice.   The evolution equations, in vacuum,
are
\begin{eqnarray}
\partial_t \gamma_{ij} & = &
-2 \alpha K_{ij}+\nabla_i \beta_j+\nabla_j \beta_i
\\
\partial_t K_{ij} &=&
-\nabla_i \nabla_j \alpha + \alpha \left( R_{ij}+K\
K_{ij}
-2 K_{im} K^m_j \right) + \beta^m \nabla_m
K_{ij}+K_{im} \nabla_j \beta^m+K_{mj} \nabla_i
\beta^m,
\end{eqnarray}
where $K_{ij}$ is the extrinsic curvature of the 3--dimensional time slice.
The Hamiltonian constraint equation is
\begin{equation}
R+K^2-K^{ij} K_{ij}=0
\end{equation}
and the three momentum constraint equations are
\begin{equation}
\nabla_i \left(K^{ij}-\gamma^{ij} K \right)=0.
\end{equation}
Note that in the above, $K=K_{ij} \gamma^{ij}$, $\beta_i =
\beta^j \gamma_{ij}$, and $\nabla_i$ is the covariant
three-space derivative.  As we discuss in section~\ref{initdat}, the constraint
equations are used to obtain the initial data, and the evolution equations are
used to advance the solution in time.

\subsection{Definition of Variables}
We build on earlier work of Ref.~\cite{Anninos93c,Bernstein93b} in
defining the variables used in our code.  Additional metric and
extrinsic curvature variables must be introduced to allow for the
odd-parity modes present now that the system allows for rotation.  We
define the variables used in our evolutions as follows:
\begin{mathletters}
\begin{eqnarray}
\gamma_{ij} & = &\left( \begin{array}{ccc}
\gamma_{\eta\eta} & \gamma_{\eta\theta} & \gamma_{\eta\phi}\\
\gamma_{\eta\theta} & \gamma_{\theta\theta} & \gamma_{\theta\phi}
\\
\gamma_{\eta\phi} & \gamma_{\theta\phi} & \gamma_{\phi\phi}
\end{array} \right) =  \Psi^4 \left( \begin{array}{ccc}
A & C & E \sin^2\theta \\
C & B & F \sin\theta \\
E \sin^2\theta & F \sin\theta & D \sin^2\theta
\end{array} \right)
\end{eqnarray}
and
\begin{eqnarray}
K_{ij} & = & \Psi^4 H_{ij} = \Psi^4 \left( \begin{array}{ccc}
H_A & H_C & H_E \sin^2\theta \\
H_C & H_B & H_F \sin\theta \\
H_E \sin^2\theta & H_F \sin\theta & H_D \sin^2\theta
\end{array} \right).
\end{eqnarray}
\end{mathletters} \\
In these expressions $\eta$ is a logarithmic radial coordinate, and
($\theta$,$\phi$) are the usual angular coordinates.  The relation
between $\eta$ and the standard radial coordinates used for
Schwarzschild and Kerr black holes is discussed in section
\ref{initdat}.  As in Ref.~\cite{Anninos93c}, the conformal factor
$\Psi$ is determined on the initial slice and, since we do not use it
as a dynamical variable, it remains fixed in time afterwards.  The
introduction of $\Psi$ into the extrinsic curvature variables
simplifies the evolution equations somewhat.  For the purposes of our
numerical evolution we will treat $A$, $B$, $D$, $F$, all six $H$'s,
and all three components of the shift as dynamical variables to be
evolved.  Two of the shift components are used to eliminate metric
variable $C$ (this will be discussed in section~\ref{shift}), and one
shift component is used to eliminate $E$.  The various factors of
sin$\theta$ are included in the definitions to explicitly account for
some of the behavior of the metric variables near the axis of
symmetry and the equator.

Within this coordinate system, the ADM mass and angular momentum
about
the $z$-axis are defined to be~\cite{Omurchada74}
\begin{mathletters}
\begin{eqnarray}
M_{ADM} &=& -\frac{1}{2\pi}\oint_S \nabla_a (\Psi e^{-\eta/2}) dS^a
\label{admmass} \\
P_a &=& \frac{1}{8\pi} \oint_S
\left(H_a^b-\gamma_a^b H\right) dS_b.
\label{admp}
\end{eqnarray}
\end{mathletters} \\
In terms of the variables defined in this paper these expressions yield
\begin{mathletters}
\begin{eqnarray}
 M_{ADM} & = & -\int_0^\pi e^{\eta/2} \left(
\partial_\eta \Psi-
\Psi/2 \right) \sin\theta d\theta \label{admmass2},\\
J & = & P_{\phi} = \frac{1}{4} \int_0^\pi \Psi^6 H_E \sqrt{\frac{B\, D}{A}}
\sin^3\theta d\theta
\label{angularmom}
\end{eqnarray}
\end{mathletters} \\
Because of this, the variable $H_E$ is extremely important.  It
determines whether angular momentum is present in the spacetime.
Although the ADM mass is defined strictly only at spatial infinity
$I^0$, in practice we evaluate it at the edge of the spatial grid.  As
we use a logarithmic radial coordinate $\eta$, this is in the
asymptotic regime.  While the angular momentum is, in principle, also
measured at $I^0$, the presence of the azimuthal Killing vector makes
it possible to evaluate $J$ at any radius. We compute this quantity
during our evolution and use it as a test of the accuracy of our code.
This will be discussed in a later section.

\subsection{Symmetries}
Symmetries are an important consideration in the evolution code for
the setting of boundary conditions.  This is important not only for
solving numerical elliptic equations, but also for appropriate finite
differencing of our variables.  Most of the appropriate conditions can
be derived merely by considering the behavior of metric functions near
the boundaries or the symmetry operations one can perform on a
spinning object, without appealing to the Einstein equations
themselves.

The principal symmetries are:

({\it i})
Axisymmetry.  We chose to study axisymmetric spacetimes as a first
step towards understanding general rotating black holes.  Even in the
stationary, Kerr case, rotating black holes are already an inherently
2D problem (i.e., they cannot be treated as a spherical system, as
Schwarzschild is.)  The fact that any rotating black hole should
settle down to a Kerr black hole, and thus an axisymmetric solution,
makes this a good choice for understanding the late time behavior of
any rotating black hole system.  One physical restriction imposed by
this symmetry is that angular momentum cannot be radiated (see, e.g.,
p. 297 of Ref.~\cite{Wald84}).

Axisymmetry requires that the transformation $\phi \rightarrow
\phi+{\mathrm constant}$ leaves the problem unchanged, and therefore
all variables in the problem must be independent of $\phi$.  However,
it is important to realize that $\phi \rightarrow -\phi$ is {\it not}
a symmetry of this spacetime, since performing this transformation
would amount to reversing the direction of spin of the hole.  In polar
coordinates, there is an additional consequence of axisymmetry: the
transformation $\phi \rightarrow \phi+\pi$ produces the same result as
$\theta \rightarrow -\theta$, providing a boundary condition on the
symmetry axis $\theta=0$.  These considerations require the following
variables to be symmetric across the axis: $A$, $B$, $D$, $F$ $H_A$,
$H_B$, $H_D$, $H_F$, $H_E$, $\alpha$, $\beta^\eta$, and $\beta^\phi$.
The remaining variables $C$ and $H_C$ are antisymmetric across the
axis.

({\it ii}) Equatorial Plane Symmetry.  We require the spacetime to be
identical when reflected through the equator defined by $\theta =
\pi/2$ ($z=0$).  This symmetry was not strictly necessary, but
adopting it reduces the complexity of the problem slightly.  One might
think that the transformation $z \rightarrow -z$ (or equivalently
$\theta \rightarrow \pi-\theta$) would result in a hole spinning in
the opposite direction, but it does not.  Inverting through the
equator in this fashion is actually equivalent to rotating the hole by
$\pi$ radians about the $x$ axis, then by sending $\phi \rightarrow
-\phi$.  Both of these latter operations clearly result in reversing
the sense of the hole's spin, and so performing both of them leaves
the direction of the hole's spin unchanged.  One can easily check that
the Kerr solution itself is manifestly unchanged by the operation $z
\rightarrow -z$. These considerations require the following variables
to be symmetric across the equator: $A$, $B$, $D$, $H_A$, $H_B$,
$H_D$, $H_E$, $\alpha$, $\beta^\eta$, and $\beta^\phi$.  The remaining
variables $C$, $H_C$, $H_F$, and $F$ are antisymmetric across the
equator.

There is an alternate equatorial boundary condition to consider.  In
the ``cosmic screw'' (the collision of two black-holes on axis with
equal and opposite angular momenta) the sense of the rotation of the
two holes is changed upon inversion through the equatorial plane.  The
appropriate isometry is thus $(\theta,\phi) \rightarrow
(\pi-\theta,-\phi)$. $H_E$ and $\beta^\phi$ are now antisymmetric, and
$H_F$ and $F$ are symmetric.  These conditions will
be considered in a future paper.

({\it iii}) Time/rotation symmetry.  This is a symmetry of the initial
slice.  It simplifies the initial data for the extrinsic curvature by
requiring that all but the values $H_E$ and $H_F$ be zero.  This
symmetry says that the transformation $(\phi,t) \rightarrow
(-\phi,-t)$ leaves the problem unchanged.

({\it iv}) Inversion through the throat.  Building on previous
work~\cite{Anninos93c,Bernstein93b}, we construct our spacetimes to be
inversion symmetric through the black hole ``throat''.  This
Einstein-Rosen bridge~\cite{Einstein35} construction has two
geometrically identical sheets connected smoothly at the throat,
located at $\eta = 0$.  This symmetry requires the metric to be
invariant under the transformation $\eta \rightarrow -\eta$.  For the
case of the Kerr spacetime, it can be expressed as
\begin{equation}
\bar{r} \rightarrow \left(m^2-a^2\right)/
\left(4 \bar{r} \right),
\end{equation}
where $\bar{r}$ is a generalization of the Schwarzschild isotropic
radius, $a$ is the Kerr rotation parameter, and $m$ is the mass of the
Kerr black hole.  The variable $\bar{r}$ is defined as
\begin{mathletters}
\begin{eqnarray}
\bar{r}&=&\frac{\sqrt{m^2-a^2}}{2}e^\eta\label{rbareta}
\end{eqnarray}
and is related to the usual Boyer-Lindquist~\cite{Misner73} radial coordinate
via
\begin{eqnarray}
r&=&\bar{r} \left(
1+\frac{m+a}{2 \bar{r}}\right) \left(1+\frac{m-a}{2 \bar{r}}\right).
\label{rrbar}
\end{eqnarray} \label{rbar}
\end{mathletters} \\
Note that in the Kerr spacetime the horizon, located at
$r=m+\sqrt{m^2-a^2}$, is at $\bar{r}=\sqrt{m^2-a^2}/2$ in the
$\bar{r}$ coordinates, or at $\eta=0$, just as in previous studies of
the Schwarzschild spacetime~\cite{Bernstein89}.

This symmetry is perhaps the most important because it is what makes
our spacetime a black hole spacetime.  When combined with symmetry
({\it iii}) above and the differential equation for a trapped surface
it tells us that $\eta=0$ is a trapped surface on the initial slice.
To impose this symmetry, one requires that metric variables with a
single $\eta$ index will be antisymmetric across the throat, and all
others will be symmetric across the throat.  The extrinsic curvature
variables will have the opposite symmetry on the throat as the
corresponding metric variables when the lapse is antisymmetric and the
same symmetry when the lapse is symmetric.  Thus, the following are
symmetric: $A$, $B$, $D$, $F$, $H_E$, $H_C$, $\beta^\theta$, and
$\beta^\phi$.  The following are antisymmetric: $C$, $E$, $H_A$,
$H_B$, $H_D$, $H_F$, $\beta^\eta$ (this is the symmetry of the $H$'s
when $\alpha$ is antisymmetric).

It is possible to slice the Kerr spacetime with a symmetric lapse
across the throat, but a different symmetry at the throat needs to be
employed for certain extrinsic curvature variables.  Kerr initial data
specifies that $H_E$ and $H_F$ are symmetric and antisymmetric across
the throat, respectively; Bowen and York~\cite{Bowen80} initial data
sets, discussed below, specify that $H_E$ is symmetric across the
throat.  These conditions are incompatible with an $\eta \rightarrow
-\eta$ boundary condition if a symmetric lapse is employed
($\gamma_{ij}$ must have the same symmetry as $\alpha H_{ij}$ for the
evolution equations to be consistent).  This problem could be removed
if we use $(\eta,\phi) \rightarrow (-\eta,-\phi)$ instead of $\eta
\rightarrow -\eta$, resulting in consistent evolution equations on
both sides of the throat.  We point out that this condition is
consistent with the Kerr initial data, and is the generalization of
the technique of evolving Schwarzschild with a symmetric
lapse~\cite{Bernstein93b}.
\section{Initial Data}
\label{initdat}

We can use the constraints to construct initial data.  An especially
convenient formulation of the initial data problem was given by Bowen
and York~\cite{Bowen80}.  They define a conformal spacetime in which
\begin{mathletters}
\begin{eqnarray}
\gamma_{ij} &=& \Psi^4 \hat{\gamma}_{ij} \\
K_{ij} &=& \Psi^{-2} \hat{H}_{ij}.
\end{eqnarray}
\end{mathletters} \\
In terms of the conformal extrinsic curvature, we write the $K_{ij}$ as
\begin{eqnarray}
K_{ij} & = & \Psi^{-2} \hat{H}_{ij} = \Psi^{-2} \left( \begin{array}{ccc}
\hat{H}_A & \hat{H}_C & \hat{H}_E \sin^2\theta \\
\hat{H}_C & \hat{H}_B & \hat{H}_F \sin\theta \\
\hat{H}_E \sin^2\theta & \hat{H}_F \sin\theta & \hat{H}_D \sin^2\theta
\end{array} \right).  \\
\end{eqnarray}
If the initial data set is maximally sliced (tr$K=0$), or if the value
of tr$K$ is held constant, then the conformal factor drops out of the
momentum constraint equations.  This decouples the constraints and
allows one to solve for the momentum constraints first and then to
solve the Hamiltonian constraint for the conformal factor.  For more
detail, see Ref.~\cite{York79}

The initial data sets we have developed to study distorted rotating
black holes are based on the both the ``Brill Wave plus Black Hole''
solutions of the NCSA group~\cite{Bernstein94a} and the rotating black
holes of Bowen and York~\cite{Bowen80}.  A complete discussion and
analysis of these rotating data sets will be published in
Paper III.  Here we provide only the basic construction of
the initial data sets.
Generalizing the conformally flat approach of Bowen and
York~\cite{Bowen80}, and following earlier work of
Ref.~\cite{Bernstein94a},
we write the metric with a free function $q(\eta,\theta)$:
\begin{equation}
dl^2=\Psi^4 \left[
  e^{2\left(q-q_0 \right)}
  \left(d\eta^2+d\theta^2 \right)
  +\sin^2\theta d\phi^2.
\right]
\end{equation}

The function $q_0$ is chosen so that if the function $q(\eta,\theta)$
is chosen to vanish, we are left with the Kerr 3-metric in these
coordinates, as shown below. (We note that Kerr is not conformally
flat.)  For other choices of $q(\eta,\theta)$ we obtain another
spacetime.  As we will see, appropriate choices of this function
(along with appropriate solutions to the momentum constraint for the
extrinsic curvature terms) can lead to Schwarzschild, the NCSA distorted
non-rotating black hole (as in Ref.~\cite{Bernstein94a}), the Bowen and
York rotating black hole~\cite{Bowen80}, or a distorted Bowen and York
black hole.

As in Ref.~\cite{Bernstein94a}, the function $q$, representing the
Brill wave, can be chosen somewhat arbitrarily, subject to symmetry
conditions on the throat, axis and equator, and falloff conditions at
large radii~\cite{Brandt94a,Bernstein94a}.  Often the function $q$
will be chosen to have an inversion symmetric gaussian part given by:
\begin{mathletters}
\begin{eqnarray}
q &=& \sin^n\theta\, q_{G}, \\
q_{G} & = & Q_0 \left( e^{-s_+}+e^{-s_-} \right), \\
s_\pm & = & \left(\eta \pm \eta_0 \right)^2/\sigma^2.\label{metric form}
\end{eqnarray}\label{brill}
\end{mathletters} \\
This form of the Brill wave will be characterized by several
parameters: $Q_0$ (its amplitude), $\sigma$ (its width), $\eta_0$ (its
coordinate location), and $n$, specifying its angular dependence,
which must be positive and even.  We note that Eqs.~(\ref{brill})
simply provide a convenient way to parameterize the initial data sets,
and to allow us to easily adjust the ``Brill wave'' part of the initial
data.  Many other devices are possible.

In this generalization, we may now interpret the
parameter $q_0$ as the Brill wave $q$ required to make the spacetime
conformally flat.
A distorted Bowen and York spacetime can be made by
setting $q= q_{G}+q_0$ and a distorted Kerr
spacetime can be made by setting $q=\sin^n\theta q_{G}$, when appropriate
solutions are taken for the momentum constraints.

As a consequence of this generalization of the metric, we must now solve
both the Hamiltonian and momentum constraints.  The Hamiltonian
constraint equation is
\begin{equation}
\frac{\partial^2 \Psi}{\partial\eta^2}+\frac{\partial^2
\Psi}
{\partial\theta^2}+
\frac{\partial \Psi}{\partial\theta}\cot\theta -\frac{\Psi}{4} =
-\frac{\Psi}{4} \left(
\frac{\partial^2}{\partial\eta^2}\left(q-q_0\right)+
\frac{\partial^2}{\partial\theta^2}\left(q-q_0\right)
\right)
-\frac{\Psi^{-7}}{4}\left( \hat{H}_E^2
\sin^2\theta+\hat{H}_F^2 \right),
 \label{Energy Constraint}
\end{equation}
and the $\phi$ component of the momentum constraint (the only
non-trivial momentum constraint equation for the initial data) is
\begin{equation}
\partial_\eta \hat{H}_{E} \sin^3\theta+
\partial_\theta \left( \hat{H}_F \sin^2\theta \right) =0.
\label{mom constraint}
\end{equation}

Based on these ideas, we consider several solutions to the initial
value problem for evolution in this paper.  Complete details of these
data sets and extensions of them will be discussed in
Paper III.  The first solution we consider is the Kerr
solution, where
\begin{mathletters}
\begin{eqnarray}
\Psi_0^4&=&g^{(K)}_{\phi\phi}/\sin^2\theta \label{psiset}\\
e^{-2 q_0}&=&g^{(K)}_{rr} \left( \frac{dr}{d\eta} \right)^2
=g^{(K)}_{\theta\theta}\\
\hat{H}_F &=& -2 \Psi_0^2
\alpha\,a^3\,m\,r\,\cos\theta \sin^3\theta \rho^{-4} \\
 \hat{H}_E &=& \Psi_0^2
\alpha\,a\,m \sin^2\theta \left( \rho^2 \left(r^2-a^2 \right)
	+2 r^2 \left( r^2+a^2 \right) \right) \Delta^{-1/2} \rho^{-4}
\label{q0set},\\
\rho^2 &=& r^2+a^2\, \cos^2\theta, \\
\Delta &=& r^2-2\, m\, r+a^2.
\end{eqnarray}
\end{mathletters} \\
Here $\Psi_0$ is the value of $\Psi$ for the Kerr spacetime, and
$g^{(K)}_{ij}$ denotes a Kerr metric element in Boyer-Lindquist
coordinates, as given in Ref.~\cite{Misner73}.  Although this may not
look familiar, one may check that with the transformations
\begin{mathletters}
\begin{equation}
r  = r_+ \cosh^2\left(\eta/2\right)- r_- \sinh^2\left(\eta/2\right),
\end{equation}
and
\begin{equation}
r_\pm = m \pm \sqrt{m^2-a^2}\label{eta coord transform}
\end{equation}
\end{mathletters} \\
the Kerr solution in Boyer-Lindquist coordinates actually results, as
detailed in Paper III.  Note that this is the same transformation
given in Eqs.(\ref{rbar}).  Note also that if $a=0$ then $q_0=0$ and
we recover the Schwarzschild 3-metric.  This metric is now in the form
used in the previous NCSA black hole
studies~\cite{Abrahams92a,Anninos93c,Bernstein93b,Bernstein94a,Bernstein94b}.

Another solution we study is related to the original solution of
Bowen and York~\cite{Bowen80}.  Following them, we choose
\begin{mathletters}
\begin{eqnarray}
\hat{H}_E&=&3 J \\
\hat{H}_F&=&0
\end{eqnarray}
\end{mathletters}  \label{bymom}\\
where $J=a\,m$ is the
total angular momentum of the spacetime. But then we may distort it by
choosing
\begin{equation}
q = q_G \sin^n\theta+q_0 \label{qdef}
\end{equation}
and solving the Hamiltonian constraint.

A third solution is the odd-parity distorted non-rotating black hole, which
we create by choosing
\begin{mathletters}
\begin{eqnarray}
\hat{H}_E & = & q_G^\prime \left(\left(n^\prime+1\right)-
\left(2+n^\prime \right) \sin^2\theta \right) \sin^{n^\prime-3}\theta \\
\hat{H}_F & = & -\partial_\eta q_G^\prime
\cos\theta \sin^{n^\prime-1} \theta, \label{odd parity solution}
\end{eqnarray}
\end{mathletters} \\
which has ``odd-parity'' radiation but no rotation, as one can verify
by checking Eq.~(\ref{angularmom}).  Because $H_E$ vanishes in the
large $\eta$ limit, the ADM integral for the angular momentum must
also vanish (See Eq.(\ref{angularmom})).  Also note that $n^\prime$
must be odd and have a value of at least 3.

Thus, in some sense this third solution is a distortion from a
Schwarzschild black hole, but one that is fundamentally different from
the ``Brill wave plus black hole'' data sets of
Ref.~\cite{Bernstein93b}. This data set describes a non-rotating
spacetime, parts of which are ``twisting and untwisting,'' but without
any angular momentum.  Some of these data sets will be evolved and
studied in detail in Paper II.

For all these data sets we need to solve the nonlinear elliptic
Hamiltonian constraint equation.  We accomplish this by replacing
$\Psi$ with $\Psi+\delta \Psi$ and expanding in $\delta \Psi$ to
create a linear elliptic equation and iteratively solving for
$\delta \Psi$.  The linear elliptic equation is solved by employing
finite differencing and the multigrid elliptic solver described in
section~\ref{gauge}. For a more detailed discussion, see Paper III.

\section{The Evolution Code}
\label{code}
In this section we discuss our evolution code, including various gauge
choices and computational details.

\subsection{Gauge Choices}
\label{gauge}
\subsubsection{Lapse}
\label{Lapse}
Maximal slicing is used, with a lapse that is antisymmetric across
the throat.  Maximal slicing is the condition that the trace of the
extrinsic curvature is zero throughout the evolution.
We note that the Kerr solution in standard form is maximally sliced
with antisymmetric lapse.
Setting tr$K=0$ in the evolution for tr$K$ gives
\begin{equation}
0=-\nabla^a \nabla_a \alpha+\alpha R.\label{trkz}
\end{equation}
We find that the code is more stable if we use the energy constraint
equation to eliminate the Ricci scalar $R$ from Eq.(\ref{trkz}) since
it contains second derivatives of the metric functions.  The behavior
of second derivatives can be troublesome near the peaks in metric
functions that can develop near black holes.  The evolution equation
for tr$K$ becomes, therefore,
\begin{equation}
0=-\nabla^a \nabla_a \alpha+\alpha K^{ij} K_{ij}.
\label{lapse eqn}
\end{equation}
This equation is solved numerically on each time slice during the
evolution using a multigrid solver to be discussed in section
\ref{Compute}.

Maximal slicing is an important slicing condition for its singularity
avoiding properties.  As the black hole evolves, constant $\eta$
observers are really falling into the hole.  Observers at the same
value of $\eta$ but different values of $\phi$ along the equator will,
for example, find that they are closer together as the evolution
proceeds.  If we were not careful, our code would follow these
observers as they fall in until the distance between them becomes zero
(as they reach the singularity) causing our code to crash.  For
Schwarzschild and distorted non-rotating spacetimes maximal slicing
prevents the evolution from reaching the singularity.  Instead, when
the slice approaches the radius $r=\frac{3}{2}M$ the lapse goes to
zero and proper time will not advance at that
location~\cite{Estabrook73}.

For rotating black holes the singularity structure inside the horizon
is significantly different.  For these reasons one might worry that
the singularity avoidance properties could be different.  The interior
region of a rotating black hole has a timelike singularity, Cauchy
horizons, and an infinite patchwork of ``other universes''
\cite{Carter65}.  However, this problem was investigated by Eardley
and Smarr~\cite{Eardley79}, who showed that maximal slicing does in
fact avoid the singularity inside the Kerr black hole (in fact, it
avoids the Cauchy horizon), although in a different manner from
Schwarzschild.  They estimated the location of a limit surface, which
was found with more precision by Duncan~\cite{Duncan85} who obtained
\begin{equation}
r_{lim} \approx 3 M \left[ 1 + \left( 1-8 a^2/9 M^2 \right)^{1/2} \right]/4,
\end{equation}
where $r_{lim}$ is given in terms of the standard Boyer-Lindquist
coordinates.  However, we will measure ``the circumferential radius,''
not the Boyer-Lindquist $r$ coordinate, so the above equation must be
converted using the following formula:
\begin{equation}
r_c = \sqrt{r^2+a^2+2 a^2 m/r}.
\end{equation}
This is simply the result of evaluating $r_c = \Psi^2 \sqrt{D}$ at
$\theta=\pi/2$ for the analytic Kerr metric.

Thus the limit surface depends on the rotation parameter.  Note that
if the rotation parameter $a$ vanishes, we recover the well known
limiting surface $r_c=3M/2$ for Schwarzschild.  Actual time slices
will approach but do not fall inside this radius interior to the
horizon.  However, we usually use an antisymmetric slicing condition
at the throat, causing the lapse to vanish there, preventing $r_c$
from evolving on throat.  Further out from the throat, however, the
coordinate system will evolve toward this limiting surface.

We have tested these ideas by numerically evolving rotating black
holes with maximal slicing using our code, described in detail below.
First we evolve a non-rotating hole, but with our new black hole
construction described above.  In Fig.~\ref{fig:limit,o1} we plot the
limit surface using the circumferential radius for an odd-parity
distorted non-rotating hole, and we see that it locks onto the
familiar $r_c=\frac{3}{2}M$ surface. The parameters describing the
initial data for this calculation are
$(Q_0^\prime,n^\prime,\eta_0^\prime,\sigma^\prime)=
(2.0,\,3,\,1.0,\,1.0)$.  This run is described as run {\it o1} in
Paper II.

Next, Fig.~\ref{fig:limit} shows the computed value of
$r_c$ for a rapidly rotating ($a/m=.70$) black hole.  The calculation
of $r_c$ from the data on the slice $t=60M$ is given with a solid
line, and the limit surface predicted by Duncan with a dashed line.
In terms of the parameters in section~\ref{initdat} the initial data
for the $a/m=.70$ spacetime is described by
$(Q_0,\,\eta_0,\,\sigma,\,J,\,n)=(1.0,\,1.0,\,1.0,\,10.0,\,2)$. This
same calculation is labeled run {\it r4} in Paper II, where its
evolution is analyzed in detail.

As one can see, maximal slicing does indeed serve to limit the advance
of proper time inside a rotating black hole, in spite of the fact that
the singularity structure is so different from that of a non-rotating
black hole.  The slices do converge to the correct radius, as
predicted by Duncan, near the horizon ($\eta \approx 2.7$).  Outside
this region the radius is, of course, much larger.  This spacetime is
not expected to have exactly the same limiting structure as the
symmetrically sliced Kerr spacetime, since it has been strongly
distorted by gravitational waves and since an antisymmetric lapse has
been employed (freezing the evolution at the throat); hence the
variations about Duncan's prediction.  We have tested this prediction
for limit surfaces for a number of rotating black holes and find
similar results in all cases.  We note that this observation could be
used to extract the rotation parameter, $a/m$, from the maximally
sliced spacetime.  We also note that these figures are very useful for
estimating the location of the apparent horizon, even without an
apparent horizon finder, since the slices will wrap up close to the
limit surface inside, but move quickly away from it as one moves out,
crossing the horizon.

\subsubsection{Shift}
\label{shift}
In previous work on non-rotating black holes a shift vector was chosen
to make the off-diagonal component of the 3--metric
$\gamma_{\eta\theta}$ vanish~\cite{Bernstein93b} (in that system no
other off-diagonal components were present).  This condition was found
to be crucial to suppress a numerical instability occurring near the
axis of symmetry.  An elliptic equation for the two components of the
shift ($\beta^{\eta}$ and $\beta^{\theta}$) present in that system
provided a smooth shift and stable evolution.  Following this
philosophy we have developed a shift condition that generalizes this
approach to the rotating case.

The gauge condition used in our evolutions is $C=0$ and $E=0$.  Our
choice has the property of reducing to the NCSA gauge used in the
non-rotating work as the rotation goes to zero, and reducing to the
Kerr shift as $Q_0 \rightarrow 0$.  Since the Kerr shift allows the
stationary rotating black hole metric to be manifestly time
independent, one expects that for the dynamical case a similar shift
will be helpful.  The evolution equations for $C$ and $E$ can be used
to construct a differential equation for the shifts. Other
off-diagonal terms may be eliminated through appropriate shift
choices, and as with the shift used in the non-rotating system, there
is additional gauge degree of freedom that has not been exploited.  We
note that the quasi-isotropic shift has been used successfully in a
recent study of rotating matter collapse~\cite{Abrahams94a}.

Let us now consider how to implement the condition $C=0$ and $E=0$.
First, the relevant metric evolution equations are:
\begin{mathletters}
\begin{eqnarray}
\partial_t C =&0&=-2 \alpha H_C +
A \partial_\eta \beta^\theta +B \partial_\eta \beta^\eta + F
\sin\theta \partial_\eta \beta^\phi \\
\partial_t E =&0&=-2 \alpha H_E +
D \partial_\eta \beta^\phi
+F \beta^\theta_{,\eta}/\sin\theta.\label{evoHE}
\end{eqnarray}
\end{mathletters} \\
These equations can be combined to produce a single equation involving
$\beta^{\eta}$ and $\beta^{\theta}$:
\begin{equation}
2 \alpha \left(H_C-\frac{F\sin\theta}{D} H_E \right) =
A \partial_\eta \beta^\theta+\left( B - \frac{F^2}{D} \right)
\partial_\theta \beta^\eta.
\end{equation}

We can solve this equation by introducing an auxiliary function
$\Omega$ through the definitions:
\begin{mathletters}
\begin{eqnarray}
\beta^\eta &=& \partial_\theta \Omega,\\
\beta^\theta &=& \partial_\eta \Omega,
\end{eqnarray}\label{defshift}
\end{mathletters} \\
(following Ref.~\cite{Bernstein94a}), producing an elliptic equation
for the function $\Omega$:
\begin{equation}
2 \alpha \left(H_C-\frac{F\sin\theta}{D} H_E \right) =
A \partial_\eta^2 \Omega+\left( B - \frac{F^2}{D} \right)
\partial_\theta^2 \Omega.
\end{equation}
This equation is then solved by finite differencing using a numerical
elliptic equation solver discussed in the next section.  The solution
$\Omega$ is then differentiated by centered derivatives to recover the
shift components $\beta^\eta$ and $\beta^\theta$ according to
Eqs.~(\ref{defshift}).  In practice, these shifts remain fairly small
during the evolution.  Their main function is to suppress the axis
instability, as noted in Ref.~\cite{Bernstein94a} where a similar
shift was used.

Once $\Omega$ is known, $\beta^\phi$ can be calculated by integrating
Eq. (\ref{evoHE}):
\begin{equation}
\beta^\phi=\int_{\eta_{max}}^{\eta} d\eta \left( 2 \alpha H_E
- F\partial_\eta \beta^\theta/\sin\theta \right)/D.
\end{equation}
Only one boundary condition needs to be set (the outer boundary
condition is most convenient), and it is generally set equal to the
Kerr value.  The inner boundary condition, that $\beta^\phi$ must be
symmetric across the throat, is guaranteed by Eq. (\ref{evoHE}).  This
shift component is needed to keep the coordinates from becoming
``tangled up'' as they are dragged around by the rotating hole.
Without such a shift the coordinates would rotate, leading to metric
shear~\cite{Smarr78a}.  This shift component, $\beta^\phi$, is
typically larger than $\beta^\eta$ or $\beta^\theta$.

This method for obtaining the shift has proved effective, although
there are some numerical difficulties that should be mentioned.
First, since $\beta^{\phi}$ is computed by integrating an ODE, errors
tend to accumulate as the integration progresses inward.  This can
cause trouble when integrating across the sharp peaks that develop in
the metric functions near the horizon of a black hole (See
section~\ref{metric functions} below for discussion of these peaks).
On the other hand, this occurs in the region where the lapse collapses
significantly (near the inner portion of the grid), and is not
noticeable before the axis instability sets in (see section~\ref{metric
functions} for a discussion of the instability), so it has not been a
serious problem.  For Kerr initial data, the numerically computed
shift had a maximum deviation from the analytic solution of about
$0.1\%$ on a $200 \times 55$ grid. We show a plot of the shift in
Fig.~\ref{fig:shift} for a Kerr black hole with a rotation parameter
of $a/m=.676$.  For dynamic black holes, the shift takes on a similar
form.

In addition, each $\theta=$constant line of integration is independent
of the others.  As a result of this decoupling, there can be
fluctuations in the shift across different angular zones near the
axis.  This becomes more apparent at late times when the axis
instability, common in most axisymmetric codes, sets in.  As noted
above this problem is not noticeable until after the instability sets
in, and so it is not a cause of difficulty.  An example of this
problem is shown in the next section.

\subsection{Computational Issues and Numerical Issues}
\label{Compute}
The code was developed using MathTensor, a package that runs under
Mathematica, to convert the equations to Fortran readable form.
Scripts have been written that automatically produce Fortran code
given symbolic input in MathTensor, so different variable choices
and gauge conditions can be tested fairly easily.

The actual evolution code was written in Fortran 77 to run on the Cray
Y-MP and the Cray C-90.  Currently, for a $200 \times 55$ grid it
obtains about $160$ MFlops on a single processor C-90.  About 70\% of
the time to evolve a spacetime with this code is spent solving
elliptic equations (two elliptic equations must be solved on each time
slice).  The multigrid linear system solver used by our code for the
elliptic equations was provided by Steven Schaffer of New Mexico
Technical Institute~\cite{Schaffer92} (UMGS2).  We find that although
it does not achieve the highest code performance in the traditional
sense (measured by MFlops), it produces a solution in very few
iterations, so the time to solution is often less than we achieve with
``higher performance'' iterative solvers~\cite{Towns92}.  UMGS2 is a
semi-coarsening multigrid code.  This differs from full multigrid
(described in detail in the above reference) by only performing the
coarsening along the angular dimension of the grid.  This is quite
useful because the spacetimes vary much less in the angular direction
than in the radial direction.

The evolution method that we have chosen is leapfrog.  Leapfrog is an
explicit evolution scheme that requires us to keep the metric
variables on two time steps, and the extrinsic curvature data on two
other time steps sandwiched between the metric variables.
Schematically the evolution looks like this:
\begin{mathletters}
\begin{eqnarray}
\tilde{\gamma}_{t+\Delta t/2} &=& \frac{3}{2} \gamma_t-\frac{1}{2}
\gamma_{t-\Delta t} \\
\gamma_{t+\Delta t} &=& \gamma_t+\Delta t \dot{\gamma} \left(
K_{t+\Delta t/2}, \tilde{\gamma}_{t+\Delta t/2} \right) \\
\tilde{K}_{t+\Delta t} &=& \frac{3}{2} K_{t+\Delta t/2}-\frac{1}{2}
K_{t-\Delta t/2}\\
K_{t+3 \Delta t/2} &=& K_{t+\Delta t/2}+\Delta t \dot{K} \left(
\tilde{K}_{t+\Delta t},\gamma_{t+\Delta t} \right)
\end{eqnarray}
\end{mathletters} \\
Spatial derivatives needed in the above equations were calculated
using centered, second order finite differencing.  This scheme is
essentially the same as that discussed in Ref.~\cite{Bernstein93b}.

As in most asymmetric black hole codes to date, our code has an axis
instability (See Ref.~\cite{Bernstein93b} for a detailed discussion of
the axis instability).  This instability grows worse as rotation
increases.  An effective strategy to slow the growth of this
instability is to reduce the number of angular zones, thus keeping
them farther away from the axis.  As mentioned above, we have also
used the gauge freedom in the equations to eliminate off-diagonal
elements in the 3-metric that tend to exacerbate this instability.
However, at late times, ($t\approx 70-100M$), when peaks in metric
functions become large, strong instabilities can develop, causing the
code to crash.  These next two plots show the function $B$ for the run
labeled {\it r4}, a distorted Bowen and York (rotating) black hole.
In Fig.~\ref{fig:B_comp}(a) we see the function at $t=100M$ for a grid
resolution of $150 \times 24$.  In Fig.~\ref{fig:B_comp}(b) we see the
same function for $t=70M$ with a grid resolution of $300 \times 48$
(all angular zones are shown in the figure).  The instability is
clearly visible at earlier times in the higher resolution calculation,
while the lower resolution calculation is still stable at later times.
Note that it is the region in the {\em interior} of the black hole
that develops difficulty (the horizon is located near $\eta=3$, where
the dip in the metric function $B$ occurs).

Because the radial resolution is most crucial, we often perform
calculations with resolutions of $300 \times 30$ zones, providing both
good accuracy and late time stability.  It is the angular resolution
that is most crucial in determining when the numerical instability
becomes serious.  Higher angular resolution does give more accurate
results, but also leads to instabilities at earlier times.

We have performed a series of convergence tests on our code.
Convergence was measured along the line $\theta = \pi/4$ for a number
of metric functions, and for the conformal factor.  Because we did not
have data placed along this value of $\theta$ we interpolated it from
our existing data using a third order interpolation scheme.  The
convergence rate of a given quantity was calculated by comparing
results obtained at three resolutions in a similar manner as reported
in Ref.~\cite{Bernstein93b}.  The basic principle is to assume the
error in a given quantity is proportional to $(\Delta \eta)^\sigma$,
and then the convergence rate $\sigma$ is determined experimentally.
For a formally second order accurate numerical scheme, such as ours,
one expects $\sigma \approx 2$.

For the purposes of these tests we required that $\Delta \theta =
\Delta \eta$.  As in previous work~\cite{Bernstein93b} we performed
most of the tests with a unit lapse.  We found, in general, that all
quantities converged to second order with slight variations throughout the
domain.  This applied to both high amplitude Brill wave data sets and
rapidly rotating data sets.  In Fig.~\ref{fig:converge} we show the
result of a convergence test for a pure Kerr spacetime with $J=5$
(this turns out to have $a/m=.676$).  We evolved it at three different
resolutions corresponding to $75$, $150$, and $300$ radial zones and
checked the convergence of the radial variable $A$.  The results are
shown after $4.8 M$.  In two places the direction of the convergence
changes, making the convergence almost impossible to measure there.
These points are each labeled ``crossing point'' in the figure.  These
results are typical for a variety of convergence tests we have
performed.

\section{Discussion}
\label{metric functions}

\subsection{Comparison with 1D codes}

In this section we discuss the evolution of the metric functions, both
to compare the present code to the non-rotating NCSA code, and to show
what effect rotation has on the system. For these purposes we compare
and contrast the evolution of two different black hole spacetimes
representing a wide range of the kinds of problems our code can
evolve.  The first is a pure Schwarzschild black hole, evolved with a
symmetric lapse.  This has become a classic test problem for black
hole codes~\cite{Anninos93c,Bernstein93b,Bernstein89}.  Because the
system can be evolved very accurately in 1D (given sufficient
resolution), and also because it is a difficult problem due to very
large peaks that develop in the solution~\cite{Seidel92a}, it is a
strong testbed calculation.  In Fig.~\ref{fig:1d_comp} we show a
comparison of our new code for rotating spacetimes with a 1D code
described in Ref.~\cite{Bernstein93b}.  Both codes were run with the
same radial resolution, $\Delta \eta \approx 0.020$, with the same
time steps $\Delta t=\Delta \eta$.  In the 2D case the angular
resolution was taken to be $\Delta \theta \approx 0.033$. The radial
metric function $A$ is shown at several times for both codes.  Only a
single angular zone is shown: The initial data are spherically
symmetric, and our codes maintain this symmetry to a high degree of
accuracy throughout the evolution.  All other angular zones are
indistinguishable.  The agreement between the two codes is excellent,
through the time $t=100M$ shown here.  It is important to note that
the 2D evolution was performed as a full 2D problem without
specializing it in any way.  For example, the maximal slicing equation
was solved as a full 2D elliptic equation in the 2D code, whereas it
is solved as a simple ordinary differential equation in the 1D code.

A key feature of this and the other black hole spacetimes studied here
is that constant $\eta$ observers fall inward toward the singularity,
actually passing through the horizon in finite time.  As a result,
more and more grid zones represent regions inside the horizon as the
evolution progresses.  Because of the differential rate at which these
observers fall through the horizon the grid stretches, creating a
sharp peak in the radial metric function $A$ as the evolution
continues, as shown in Fig.~\ref{fig:1d_comp}.  This effect assures us
that the code will eventually become inaccurate and crash.  This is
the bane of all black hole evolution evolution codes at present, and
is the main motivation for considering apparent horizon boundary
conditions that can, in principle, eliminate this
problem~\cite{Seidel92a,Anninos94e}.

We have also evaluated the error in the Hamiltonian constraint.  This
error is in the form of a mass density which is given by
\begin{equation}
\rho = \frac{1}{16 \pi}\left( R-K_{ab}\, K^{ab}+K^2 \right).
\end{equation}
The error in this quantity is dominated by the axis instability.
However, because this instability grows most rapidly near the horizon
its effect on the code is partially cancelled by the collapse of the
lapse.  Because we are not as interested in violations of the
Hamiltonian constraint inside the horizon, since this should not
affect what happens outside for hyperbolic evolution, we will be more
interested in the quantity $\alpha \vert \rho \vert$. (Although we do
solve elliptic equations, which propagate information instantaneously,
these equations are for {\em gauge} conditions.  An ``error'' in a
gauge condition, which is arbitrary, does not affect physical results
in principle.)  The maximum error in this quantity as a function of
time for various runs is plotted in Fig.~\ref{fig:ham_max}.  Our
evolutions are generally unable to proceed past the point where this
maximum density passes unity.  The Hamiltonian constraint violation
illustrated by the maximum error is dominated by the axis instability,
and highly localized.

Because of the strong locality of this error and the hyperbolic nature
of our evolution, we do not feel, however, that the maximum error
provides us with the best understanding of the overall accuracy of the
code.  For this reason we also consider the average value of $\vert
\rho \vert$ over the grid (weighted by the lapse function $\alpha$).
We provide a plot of this measure of the error for the same runs in
Fig.~\ref{fig:ham_avg}.  It should be noted that this quantity is
generally $4$ - $8$ orders of magnitude smaller than the maximum error
on the grid for all runs considered.  The runs that represent
distorted spacetimes are initially less smooth, and for this reason
the Hamiltonian violation, is larger than for Schwarzschild.  However,
it is small and grows at a rate comparable to that of the
Schwarzschild evolution.

\subsection{Comparison with the 2D codes}

Although the 1D problem is a good test of the longitudinal component
of the field it does not test the code's ability to handle a highly
non-spherical, distorted black hole.  The next case we consider is a
distorted, non-rotating black hole with a Brill wave, labeled run {\it
r0} and is specified by $(Q_0,\eta_0,\sigma,J,n)=(1.0,1.0,1.0,0.0,2)$.
Initially, this data set is a highly distorted and nonspherical black
hole that evolves in an extremely dynamic way, so it is a very strong
test case for this new code.  In Fig.~\ref{fig:code_comp} we show the
evolution of this distorted non-rotating black hole spacetime obtained
with our code, and compare it to the same evolution obtained with the
code described in Ref.~\cite{Anninos93c,Bernstein93b}.  The radial
metric function $A=g_{\eta\eta}/\Psi^4$ is shown at time $t=60
M$. Only a single angular zone is plotted ($\theta=\pi/2$), as by this
time the developing peak is quite spherical and nearly independent of
$\theta$.  In the figure we compare the evolution obtained with the
rotating code with symmetric lapse, antisymmetric lapse conditions,
and the non-rotating code of Ref.~\cite{Anninos93c,Bernstein93b} with
symmetric lapse.  When the same lapse conditions are used, both codes
show excellent agreement.  Note also that even with an antisymmetric
lapse (though the early evolution is not shown) the runs will be
noticeably but minimally different at late times.  This is the first
time these data sets have been evolved with this antisymmetric slicing
condition.

We have compared many other aspects of results of our code with those
obtained with the code described in
Ref.~\cite{Anninos93c,Bernstein93b} and find excellent agreement for
metric functions, extracted waveforms, horizon masses, etc.  The
comparisons have been performed for a wide range of data sets.  A
detailed study of horizons and waveforms will be presented in
Paper~II.

\subsection{Rotating spacetimes}

Rotating spacetimes differ only slightly from the picture discussed
above. Constant $\eta$ observers along the equator are, in general,
rotating about the black hole.  Because of this, their fall through
the horizon is slower and grid stretching is less near the equator.
The peak in the metric function $A$ is, therefore, lesser in magnitude
here.  This gives the data for a rotating black hole spacetime an
obvious angular dependence not present in distorted non-rotating
metric data.

The pure stationary Kerr spacetime is, in general, much less stable
numerically than the Schwarzschild, or even the distorted Kerr, and
special care is required to evolve it.  We have two methods which we
use to carry out the evolution.  The first is to use a lapse which is
symmetric across the throat.  We begin our discussion of rotating
black hole evolutions by showing a surface plot of metric variable
$A=g_{\eta\eta}/\Psi^4$ for a symmetrically-sliced high resolution
($300 \times 30$) $J=5$ ($a/m=.68$) Kerr spacetime in
Fig.~\ref{fig:A_at_80M}.  The error in $J$ as computed by
Eq.(\ref{angularmom}) reached a maximum error of $3.6\%$ over the grid
at this late time, and occurs over the peak in $g_{\eta\eta}$.  This
is an important and nontrivial test, as $J$ must be conserved during
the evolution even though the different metric and extrinsic curvature
components entering Eq.(\ref{angularmom}) evolve dramatically.

Next we consider this same spacetime with an antisymmetric lapse.  The
antisymmetric, maximal sliced Kerr spacetime is actually a time
independent analytic solution, but it is numerically unstable.  Any
slight numerical error will destroy the perfect balance between the
Ricci terms and the derivatives of the lapse in the extrinsic
curvature evolution equations.  Then the solution will evolve rapidly,
behaving much like the symmetrically sliced Kerr spacetime.  In order
to evolve this system stably for long times, we force metric variable
$F=g_{\theta\phi}/\Psi^4$ to be zero at all times (In this stationary
spacetime this must be true).  Extending the grid to larger radii is
also helpful.  When both of these things are done we are able to
evolve past $50 M$ before serious problems develop.  The $J=5$ Kerr
evolution with an antisymmetric lapse at the throat begins to look
like the symmetric lapse run at late times.  A peak forms in the
radial metric function $A$, although a much smaller one than is seen
with the symmetric lapse.  This is illustrated in
Fig.~\ref{fig:A_at_50M}.  Because the lapse is collapsed in the inner
region $\alpha$ is virtually zero over a number of grid zones and this
satisfies both the symmetric and antisymmetric conditions to a good
approximation.  Note that even the Schwarzschild spacetime is
difficult to evolve with an antisymmetric lapse since the axis
instability becomes quite serious at about $60M$.  The error in $J$ as
computed by Eq.~(\ref{angularmom}) reached a maximum error of $1.6\%$
over the grid.

Finally, we present a distorted rotating black hole.  Calculation
labeled run {\it r4} is specified by
$(Q_0,\eta_0,\sigma,J,n)=(1.0,\,1.0,\,1.0,\,10.0,\,2)$.  These labels
correspond to the same simulations discussed in detail in Paper~II,
where many physical properties of the spacetimes are analyzed.  On a
$300 \times 48$ grid, the maximum error in the momentum integral was
$1.6\%$. Again, at this high angular resolution the axis instability
develops rapidly after about $t=70 M$, causing the code to crash.
Lower resolution runs can be carried out past $t=100 M$ with similar
results.

In Fig.~\ref{fig:A_at_60M} we show a surface plot of the radial metric
function $A$ for the rotating hole at time $t=60 M$.  Although the
familiar peak develops in this function, it does develop the expected
angular dependence.  This is typical of all our rotating black hole
evolutions.

Another typical feature of these spacetimes is ``slice-wrapping,''
discussed in section~\ref{Lapse}.  This name refers to the fact that
slices inside the horizon approach a limit surface with roughly
constant $r$ value.  This is connected to the discussion of the limit
surfaces above, but here we use this feature to illustrate the full
2D behavior of the metric functions to show that they
evolve as expected.  As a result of slice-wrapping, in our spacetimes
the value of $\Psi^2 \sqrt{D}$ (which is the circumferential radius
when evaluated at $\theta=\pi/2$) becomes constant over larger regions
of the spacetime as the evolution continues.  In a non-rotating
spacetime with a lapse that is symmetric across throat this value goes
to $3 M/2$.  These spacetimes have a different limiting value, as
discussed in section~\ref{gauge}.
In Fig.~\ref{fig:slicewrap} we show
an example of this effect for the spacetime labeled r4.  The
slice-wrapping effect is clear inside the radius of about $\eta
\approx 2.7$ where the slice moves out away from the limit surface,
out across the horizon.  From studies of horizons of black holes like
those in Ref.~\cite{Anninos93a}, we know the horizon is located near
$\eta=2.7$.

The strong internal structure evident in $D=g_{\phi\phi}/\Psi^4$ well
inside the black hole, near the throat ($\eta=0$), is a remnant of the
initial Brill wave, indicating that this black hole was initially
quite distorted.  This structure developed early on as gravitational
waves propagated into the hole, but was ``frozen in'' as the lapse
collapsed rapidly in this region of the spacetime.

A similar behavior is observed in the metric variable
$B=g_{\theta\theta}/\Psi^4$ as in $D$, except that while $D=1$
initially, $B=e^{2 \left(q-q_0\right)}$ and so the antisymmetric
lapse ``freezes in'' these different functions at the throat.  Because
the spacetime can evolve even a short distance away from there a sharp
peak develops at the throat with this slicing condition.  Fortunately,
the lapse is always small in this region and the code does not suffer
as a result.  In Fig.~\ref{fig:B_at_60M} we show the behavior of $B$,
and Fig.~\ref{fig:D_at_60M} we show the behavior of $D$.  In both
plots we are looking at run labeled {\it r4} at time $60M$.  At high
resolution ($300 \times 48$) the axis instability becomes serious at
about $70M$, but with lower resolutions it can go past $100M$.

\section{Conclusions}
\label{Conclusions}

The study of rotating spacetimes presents a new level of complexity to
the distorted black hole, and we have developed a new code to evolve
such spacetimes by building on previous non-rotating
work~\cite{Bernstein93b}.  Although the rotation requires the
introduction of new metric and shift variables, we have been able to
bring many of the same numerical methods to bear on the problem that
have been used before.  We have shown that the new code is able to
reproduce results of Ref.~\cite{Abrahams92a,Anninos93c,Bernstein93b}
for spherical and highly distorted non-rotating holes, including the
behavior of both the metric functions and derived quantities.  We have
also shown the effect of rotation on the metric functions and how they
behave differently for rotating black hole spacetimes.  In addition to
reproducing previous results, the code has also passed other tests,
such as convergence tests and the conservation of angular momentum,
and it confirms the theoretical predictions for the relation between
the limit slice for a maximally sliced black hole and its rotation
parameter.

We have also introduced a new family of distorted black hole data
sets, including distorted rotating black holes and odd-parity
distorted non-rotating black holes.  The next paper in this series
will discuss results obtained from applying this code to these new
data sets.  Specifically it will show how to calculate the location of
the apparent horizon and how to extract the waveforms for the various
$\ell$ modes radiating from the black hole.  It will also analyze the
behavior of these aspects of the spacetimes for a series of evolutions
forming a sequence of black holes with increasing rotation, and also
for odd-parity distorted black holes.

\section{Acknowledgements}
We would like to thank Andrew Abrahams, Pete Anninos, David Bernstein,
Karen Camarda, Greg Cook, David Hobill, John Jaynes, Peter Leppik,
Larry Smarr, and Wai-Mo Suen for many helpful suggestions throughout
the course of this work.  This work as supported by NCSA, and by
grants NSF PHY94-07882 and ASC/PHY93-18152 (arpa supplemented).  The
calculations were performed at NCSA on the Cray Y-MP and at the
Pittsburgh Supercomputing Center on the Cray C-90.


\begin{figure}
\caption{
This plot shows $r_c$, the circumferential radius of the maximally
sliced black hole as a function of radial coordinate $\eta$, for run
labeled {\it o1} at time $t=70 M$.  The theoretical limit surface for
a non-rotating hole is $r=1.5 M$, but our boundary condition on the
lapse prevents us from reaching that point at the throat ($\eta=0$).
Nevertheless, away from the throat, the predicted limit surface is
reached.}
\label{fig:limit,o1}
\end{figure}

\begin{figure}
\caption{
In this figure the circumferential radius $r_c$ is plotted at time $t=
60M$.  For Kerr black holes the limit surface is not $r_c=1.5 M$ but a
higher value that depends on the rotation parameter $a/m$, as
predicted by Duncan.  The dashed line is the theoretically predicted
$r={\mathrm constant}$ limit.  In regions away from the throat (where
the lapse is zero and the spacetime may not evolve) we see that this
limit surface is reached.}
\label{fig:limit}
\end{figure}

\begin{figure}
\caption{We show a 2D plot of the shift $\beta^\phi$ as computed by the
rotating code for a Kerr black hole with $J=5$.  It is virtually
identical to the analytic shift function one would obtain from the
exact Kerr solution.}
\label{fig:shift}
\end{figure}

\begin{figure}
\caption{
In both plots that follow, all of the angular zones are displayed.
(a) This 2D plot shows the contents of metric variable
$B=g_{\theta\theta}/\Psi^4$ at a resolution of $150 \times 24$ for run
{\it r4}, a highly distorted rotating black hole.  Using this
resolution, we can evolve the spacetime to $100M$ before numerical
instabilities develop.  (b) This 2D plot shows the metric variable
$B=g_{\theta\theta}/\Psi^4$ (where $\Psi$ is a conformal factor) in
run {\it r4}. It has a higher spatial resolution ($300 \times 48$) and
it develops trouble at an earlier time.  The ridges near the throat
are a result of the axis instability and the radial integration we use
to compute the $\beta^\phi$ component of the shift.}
\label{fig:B_comp}
\end{figure}

\begin{figure}
\caption{ This figure plots the convergence exponent
$\sigma$ as a function of radial coordinate $\eta$ for the radial
metric variable $A=g_{\eta\eta}/\Psi^4$ (where $\Psi$ is a conformal
factor) of a Kerr black hole spacetime with $J=5$, as measured along
$\theta=\pi/4$, evolved to $t=4.8 M$.  As discussed in the text,
$\sigma$ was measured by evolving the spacetime at 3 different
solutions.  As our methods are second order, we expect to see roughly
a horizontal line at $\sigma=2$ ($\sigma$ is the order of
convergence).  The points labeled ``crossing point'' are places at
which the direction of the convergence is changing, and we do not
expect to be able to measure the convergence well at these points.}
\label{fig:converge}
\end{figure}

\begin{figure}
\caption{
This plot compares a 1D code to our 2D code for a Schwarzschild
spacetime. Each solid (dashed) line represents the metric variable
$A=g_{\eta\eta}/\Psi^4$ (where $\eta$ is a logarithmic radial
coordinate and $\Psi$ is a conformal factor) from the 2D (1D) code at
a time interval of $10M$, beginning with $10M$ and continuing to
$100M$.}
\label{fig:1d_comp}
\end{figure}

\begin{figure}
\caption{
In this plot we show the maximum violation in the Hamiltonian
constraint function $\alpha \vert \rho \vert$ on the grid.  The error
is dominated by the effects of the axis instability and is highly
localized there, and evolution becomes impossible shortly after this
error exceeds unity.  The error is measured in units of
$M_{ADM}^{-2}$.}
\label{fig:ham_max}
\end{figure}

\begin{figure}
\caption{
In this plot we show the average of the Hamiltonian constraint
function $\vert \rho \vert$ (weighted by the lapse function).  The
errors for distorted black holes is noticeably larger since these
functions are initially less smooth.  The error is measured in units
of $M_{ADM}^{-2}$.}
\label{fig:ham_avg}
\end{figure}

\begin{figure}
\caption{In this figure we compare several calculations of the radial metric
function $A$ at time $t=60 M$ for run {\it r0} (a distorted black hole
with no angular momentum nor odd-parity distortion).  Three results
are shown: both codes were run with a symmetric lapse across the
throat, and the present rotating code was also run with an
antisymmetric lapse across the throat.  We see that the rotating and
non-rotating code produce practically identical results for the
symmetric lapse.  We also see that the antisymmetric lapse produces
similar results, except near the throat where it must be different.}
\label{fig:code_comp}
\end{figure}

\begin{figure}
\caption{This surface plot shows the radial metric function
$A=g_{\eta\eta}/\Psi^4$ (where $\eta$ is a logarithmic radial
coordinate and $\Psi$ is a conformal factor) at time $t=80M$ for an
antisymmetrically sliced Kerr spacetime with $J=5$ ($a/m=.68$).
The characteristic peak in $A$ is well-developed and its angular
dependence, explained in the text, is clearly visible. }
\label{fig:A_at_80M}
\end{figure}

\begin{figure}
\caption{This surface plot shows the radial metric function
$A=g_{\eta\eta}/\Psi^4$ (where $\eta$ is a logarithmic radial
coordinate and $\Psi$ is a conformal factor) at time $t=50M$ for an
antisymmetrically sliced Kerr spacetime with $J=5$ ($a/m=.68$).  At
this high resolution ($300 \times 30$), the axis instability sets in
shortly after this time.}
\label{fig:A_at_50M}
\end{figure}

\begin{figure}
\caption{This is a surface plot of metric function
$A=g_{\eta\eta}/\Psi^4$ (where $\eta$ is a logarithmic radial
coordinate and $\Psi$ is a conformal factor) at time $t=60M$ for the
distorted rotating black hole run labeled {\it r4}.  The presence of a
$\phi$ shift in the evolution means $\eta={\mathrm constant}$
observers near the equator feel a ``centripetal force'' and fall in
more slowly.  This results in less grid-stretching, and a lower peak
there.}
\label{fig:A_at_60M}
\end{figure}

\begin{figure}
\caption{
This surface plot is of the function $R=\Psi^2 D^{1/2} $ for the run
labeled {\it r4}.  Along the equator this value is the equatorial
circumferential radius. This plot shows that the ``slice-wrapping''
effect affects a large portion of the grid.}
\label{fig:slicewrap}
\end{figure}

\begin{figure}
\caption{This surface plot depicts metric variable
$B=g_{\theta\theta}/\Psi^4$ (where $\Psi$ is a conformal factor) at
time $70 M$ for the run labeled {\it r4}.  Note that the angular shape
of the Brill wave on the initial data slice is preserved at the
throat, but outside this region it becomes very close in value to
metric variable $D=g_{\phi\phi}/\Psi^4$ .}
\label{fig:B_at_60M}
\end{figure}

\begin{figure}
\caption{This surface plot depicts metric variable
$D=g_{\phi\phi}/\Psi^4$ (where $\Psi$ is a conformal factor) at time
$60 M$ for the run labeled {\it r4}. Note that at the throat
$D=g_{\phi\phi}/\Psi^4$ has value $1$, as it does in the initial data.
Farther from the throat it has the same value as
$B=g_{\theta\theta}/\Psi^4$.}
\label{fig:D_at_60M}
\end{figure}

\end{document}